\documentclass{PoS}

\usepackage{amsmath} 
\usepackage{amssymb} 
\usepackage{slashed}
\usepackage{braket}
\usepackage{enumerate}
\usepackage{graphics}
\usepackage{graphicx}

\title{Correlation functions with Karsten-Wilczek fermions}

\ShortTitle{Correlation functions with Karsten-Wilczek fermions}

\author{\speaker{Johannes Heinrich Weber}\\
        Technische Universit\"{a}t M\"{u}nchen, Physik Department T30f\\
        E-mail: \email{johannes.weber@tum.de}}

\abstract{
The Karsten-Wilczek action describes two chiral fermions but breaks the symmetries under both charge conjugation ($ \widehat{C}) $ and time reflection ($ \widehat{\Theta} $) explicitly, though invariance under~$ \widehat{C\Theta} $ and a mirror fermion symmetry~$ \widehat{T\Theta} $ are maintained. 
These proceedings outline how the action's symmetries and the presence of the second fermion emerge in mesonic correlation functions. 

The residual symmetries explain the non-observation of broken time-reflection symmetry in a class of mesonic correlation functions. Time-reflection symmetry is enforced for correlation functions that are manifestly invariant under $ \widehat{C} $ or $ \widehat{T} $. 
Due to contributions from the second fermion, oscillating contributions arise in some mesonic correlation functions. A second condition for non-perturbative tuning of the relevant counterterm is obtained from these oscillations. Both non-perturbative tuning conditions are independent and agree within errors.
Due to contributions from the second fermion, additional pseudoscalar states are observed in non-standard channels. Mass splittings between these additional states and the Goldstone boson vanish as $ \mathcal{O}(a^2) $.

}

\FullConference{The 32nd International Symposium on Lattice Field Theory,\\
		23-28 June, 2014\\
		Columbia University New York, NY}

\begin{document}

\section{Introduction}

Minimally doubled fermions are a category of ultralocal, chiral lattice fermions that have only two degenerate flavours of quarks. Karsten-Wilczek fermions~\cite{kw} are a type of minimally doubled fermions that retains two original poles of the na\"{i}ve Dirac operator on the Euclidean time axis. Since the two poles of the Karsten-Wilczek operator lie on a line in the Brillouin zone, the hypercubic symmetry is inevitably broken and further anisotropic terms are generated dynamically. \newline
 The tree-level Karsten-Wilczek fermion action reads (with Wilczek parameter $ \zeta $, $ 4\zeta^2>1$) %\small
 \begin{align}
   S^f_{+\zeta}[\psi,\bar{\psi},U] = a^4\sum\limits_{n\in\Lambda}&\ \sum\limits_{\mu=0}^3
   \bar{\psi}_n \frac{1}{2a} \gamma^\mu \left(U^\mu_n \psi_{n+\hat{\mu}} - U^{\mu\dagger}_{n-\hat{\mu}} \psi_{n-\hat{\mu}}\right) +m_0\bar{\psi}_n \psi_n \nonumber \\
    +& \sum\limits_{j=1}^3
   \bar{\psi}_n \frac{i\zeta}{2a} \gamma^0 \left(2\psi_n - U^j_n \psi_{n+\hat{j}} - U^{j\dagger}_{n-\hat{j}} \psi_{n-\hat{j}}\right) 
   \label{eq: KW tl f action}
 \end{align}\normalsize
 and the gauge action is assumed to be Wilson's plaquette action. The standard discrete symmetries of this action were reviewed first in~\cite{bbtw} and include $ \gamma^5 $ hermiticity, cubic symmetry under spatial rotations, symmetry under spatial reflections and symmetey under the product of charge conjugation~($ \widehat{C} $) and time reflection~($ \widehat{\Theta} $). \mbox{Eq.}~(\ref{eq: KW tl f action}) varies under charge conjugation ($ \widehat{C} $) or time reflection~($ \widehat{\Theta} $), since each transformation flips the sign of the Karsten-Wilczek term [lower line in \mbox{eq.}~(\ref{eq: KW tl f action})] relative to the na\"{i}ve term. This change of the action can be parameterised as a change of the Wilczek parameter as $ S^{f}_{+\zeta} \to S^{f}_{-\zeta} $. However,~$ \widehat{C\Theta} $ together is a symmetry transform. 
 Moreover, the action is invariant~\cite{ccww} under local vector and axial transforms (the latter only for $ m_0=0 $) %\small
 \begin{align}
   \psi_n \to e^{+i \alpha^V_n} \psi_n,\ \bar{\psi}_n \to \bar{\psi}_n e^{-i \alpha^V_n},& \qquad
   \psi_n \to e^{i \alpha^A_n \gamma^5} \psi_n,\ \bar{\psi}_n \to \bar{\psi}_n e^{+i \alpha^A_n \gamma^5}
   \label{eq: axial transform}.
 \end{align}\normalsize
 There is one further unitary transform $ \widehat{T} $ that interchanges the poles of the Dirac operator, %\small
 \begin{equation}
   \psi_n \to \widehat{T}\psi_n = T_n \psi_n, \quad \bar{\psi}_n \to (\widehat{T}\bar{\psi}_n) = \bar{\psi}_n T_n, \quad T_n \equiv i \gamma^0 \gamma^5 (-1)^{n_0},
   \label{eq: unitary transform T}
 \end{equation}\normalsize
 which leaves the na\"{i}ve term in the upper line of \mbox{eq.}~(\ref{eq: KW tl f action}) invariant but flips the sign of the Karsten-Wilczek term. It is related to the remnant of the discrete subgroup of the $ U(4) $-symmetry of the na\"{i}ve Dirac action (\mbox{cf.}~\cite{ks}) and permits two further symmetry transforms $ \widehat{T\,C} $ and $ \widehat{T\,\Theta} $. The latter has been pointed out as ``mirror fermion symmetry'' in~\cite{per}.  Hence, the action has one further independent non-standard discrete symmetry.
 The action in \mbox{eq.}~(\ref{eq: KW tl f action}) requires inclusion of one relevant and two marginal counterterms that were calculated perturbatively in~\cite{ccww} at one-loop level, %\small
 \begin{align}
   S^3_{+\zeta}[\psi,\bar{\psi}] =&\ \mathrm{sign}({\zeta})\, c(g_0^2)\ a^4\sum\limits_{n\in\Lambda} \bar{\psi}_n \frac{i}{a} \gamma^0 \psi_n 
   \label{eq: relevant term} \\
   S^{4f}[\psi,\bar{\psi},U] =&\ d(g_0^2)\ a^4\sum\limits_{n\in\Lambda} \bar{\psi}_n \frac{1}{2a} \gamma^0 \left(U^0_n \psi_{n+\hat{0}} - U^{0\dagger}_{n-\hat{0}} \psi_{n-\hat{0}}\right), 
   \label{eq: marginal fermion term} \\
   S^{4g}[U] =&\ d_p(g_0^2) \frac{\beta}{3} \sum\limits_{n\in\Lambda} \sum\limits_{j=1}^3 \mathrm{Re} \mathrm{Tr} (1-U^{j 0}_n),
   \label{eq: marginal gauge term}
 \end{align} \normalsize
 where $ U^{j 0}_n $ is the temporal plaquette at site $ n $. In particular, though \mbox{eq.}~(\ref{eq: relevant term}) flips its sign under $ \widehat{C} $, $ \widehat{\Theta} $ and $ \widehat{T} $, both \mbox{eqs.}~(\ref{eq: marginal fermion term}) and~(\ref{eq: marginal gauge term}) are invariant under~$ \widehat{C} $, $ \widehat{\Theta} $ and $ \widehat{T} $. The one-loop coefficients~\cite{ccww} read %\small
 \begin{align}
   c^{1L}(g_0^2) = -29.5320\, C_F\, G, \
   d^{1L}(g_0^2)= -0.125540\, C_f\, G, \
   d_{P}^{1L}(g_0^2) = -12.69766\, C_2\, G,
   \label{eq: 1L results}
 \end{align}\normalsize
 where $ G=g_0^2/(16\pi^2) $. A non-perturbative tuning scheme that fixes~$ c(g_0^2) $ as the value $ c_M $ that has minimal anisotropy of the pseudoscalar mass at finite lattice spacing was presented in~\cite{wcw}. Because no method for non-perturbative tuning of $ d(g_0^2) $ is known at present due to lack of sensitivity, its perturbative estimate is used.

\section{Higher order operators}
\label{sec: Higher order operators}

 Higher order operators are restricted by $ \widehat{C\Theta} $~ and chiral symmetries in particular. All possible $ \gamma^5 $~hermitian continuum operators of dimension five that respect chiral symmetry unless they vanish in the chiral limit are collected in table~\ref{tab: dim 5 cont ops}. They are referred to as $ \mathcal{O}_{RC} $ in the ensuing discussion, where $ R $ and $ C $ are row and column indices. Derivatives are symmetrised as $ D^\mu = \tfrac{1}{2}(D^\mu_R -D^\mu_L) $.
 
\begin{table}[hbt]
%   \small
 \begin{center}
  \begin{tabular}{|c|c|c|c|c|}
  \hline
   $ \bar{\psi}\ m_0^2 \psi $ 
 & $ \bar{\psi}\ m_0 \slashed{D} \psi $ 
 & $ \bar{\psi}\ m_0 \gamma^0 D^0 \psi $
 & $ m_0 F^{\mu\nu} F^{\mu\nu} $
 & $ m_0 F^{i0} F^{i0} $ \\
 \hline
   $ \bar{\psi}\ i m_0 \sum\limits_{j=1}^3\Sigma^{0j} D^j \psi $
 & $ \bar{\psi}\ i \sum\limits_{j=1}^3 \gamma^j F^{0j} \psi $
 & $ \bar{\psi}\ i\, \{ D^0, \slashed{D} \} \psi $ 
 & $ \bar{\psi}\ i\,m_0 D^0 \psi $
 & $ \bar{\psi}\ i\,m_0^2 \gamma^0 \psi $
  \\
 \hline
 $ \bar{\psi}\ i \gamma^0 \sum\limits_{j=1}^3 D^j D^j \psi $
 & $ \bar{\psi}\ i \gamma^0 D^0 D^0 \psi $
 & $ \bar{\psi}\ i \gamma^0\!\! \sum\limits_{j<k =1}^3\!\! \Sigma^{jk} F^{jk} \psi $
 & \multicolumn{2}{c|}{$ \bar{\psi}\ i ( \gamma^0 \slashed{D} \slashed{D} +  \slashed{D} \slashed{D} \gamma^0 ) \psi $}
  \\
 \hline
   $ F^{k j} (D^k F^{0 j}) $
 & $ F^{0 j} (D^0 F^{0 j}) $
 & $ *F^{j 0} (D^0 F^{j 0}) $
 & \multicolumn{2}{c|}{$ *F^{j k} (D^0 F^{j k}) $} \\
  \hline
  \end{tabular}
 \end{center}
 \caption{Eighteen continuum operators of dimension five can be constructed if odd numbers of the Euclidean index `$ 0 $' are allowed and $ \gamma^5 $ hermiticity is required. Chiral symmetry is required only in the chiral limit.}
 \label{tab: dim 5 cont ops}
 \vspace{-12pt}
\end{table}

\begin{enumerate}[a)]
\itemsep0em
  \item Chiral actions are invariant under the axial transform of \mbox{eq.}~(\ref{eq: axial transform}) and a simultaneous flip of the sign of $ m_0 $. This property prohibits all $ \mathcal{O}(am_0) $~corrections as in $ \mathcal{O}_{11} $, $ \ldots $, $ \mathcal{O}_{15} $. 
  \item Sums are completed by adding zero as \mbox{$ \sum\limits_{j=1}^3\Sigma^{0j} D^j = \tfrac{i}{2}[\gamma^0, \slashed{D}] $} and \mbox{$ \{\slashed{D},[\gamma^0, \slashed{D}]\} = 2 \vec{\gamma} \cdot [D^0,\vec{D}] $}. Then $ 2\mathcal{O}_{21}= \mathcal{O}_{22} + \mathcal{O}(a) $ and $ \mathcal{O}_{21} = 0+\mathcal{O}(a) $ are derived from field equations.
  \item Completion of sums and field equations yield $ \mathcal{O}_{23} =  2\mathcal{O}_{25} + \mathcal{O}(a) $ and  $ \mathcal{O}_{24}= -\mathcal{O}_{25} + \mathcal{O}(a) $. $ \mathcal{O}_{23} $~and $ \mathcal{O}_{24} $ are eliminated and $ \mathcal{O}_{25} $ is considered as $ \mathcal{O}(a^2m_0^2) $~chiral correction to $ c $ of \mbox{eq.}~(\ref{eq: relevant term}).
  \item $ \mathcal{O}_{31} $ is the Karsten-Wilczek term's continuum form. Though $ 2 (\mathcal{O}_{31} +\mathcal{O}_{32}) + \mathcal{O}_{33}  = \mathcal{O}_{34} $ is an exact relation, no operator is eliminated as the coefficient $ \zeta $ of $ \mathcal{O}_{31} $ is fixed (Wilczek parameter). 
  \item Iterative application of the field equation yields $ \mathcal{O}_{34} = 2\mathcal{O}_{25} + \mathcal{O}(a) $ and eliminates $ \mathcal{O}_{34} $.
  \item As $ \mathcal{O}_{41} $ and $ \mathcal{O}_{42} $ lack invariance under $ \widehat{\Theta} $ but are $ \widehat{C} $~and $ \widehat{T} $~invariant and $ \mathcal{O}_{43} $ and $ \mathcal{O}_{44} $ lack invariance under parity, they are prohibited. The sum $ \mathcal{O}_{43}+\mathcal{O}_{44} $ is a total derivative.
\end{enumerate}
Hence, $ \mathcal{O}_{32} $ and $ \mathcal{O}_{33} $ form a complete set of independent additional operators.
A lattice operator for~$ \mathcal{O}_{33} $, which couples an axial current to the colour magnetic field, can be constructed using $ \widehat{F}^{kl}_n $ as in the clover term for Wilson fermions. 
 Invariance under $ \widehat{TC} $ requires next-to-next neighbour terms in~$ \mathcal{O}_{32} $. As a next-neighbour operator, it would vanish at only one of the two poles of the Dirac operator. The two additional independent dimension five lattice operators are chosen as: \small
 \begin{align}
  \mathcal{O}_{32}:\quad S^{5 t} =&\ c_t a^4 \sum\limits_{n\in\Lambda} \bar{\psi}_n \frac{i}{4a} \gamma^0 \left( 2\psi_{n} - U^0_n U^0_{n+\hat{0}}  \psi_{n+2\hat{0}} - U^{0\dagger}_{n-\hat{0}}U^{0\dagger}_{n-2\hat{0}} \psi_{n-2\hat{0}}\right),  \\
  \mathcal{O}_{33}:\quad S^{5 B} =&\ c_B a^4 \sum\limits_{n\in\Lambda} \bar{\psi}_n 
  \sum\limits_{j,k,l=1}^3 \epsilon^{jkl} \gamma^5\gamma^j \widehat{F}^{kl}_n \psi_{n}.
 \end{align}\normalsize
 The gauge action is anisotropic due to \mbox{eq.}~(\ref{eq: marginal gauge term}) but does not include $ \mathcal{O}(a) $ operators and retains separate invariance under $ \widehat{C} $, $ \widehat{\Theta} $ and $ \widehat{T} $. This invariance holds for any observables without valence quarks even in full QCD, because the fermion determinant is invariant under each of the symmetries. This statement is proved in section~\ref{sec: Symmetry of the fermion determinant}. %\newline 
 Each dimension five operator varies under $ \widehat{C} $, $ \widehat{\Theta} $ and $ \widehat{T} $, but is still invariant under $ \widehat{C\Theta} $, $ \widehat{T\,C} $ and $ \widehat{T\,\Theta} $. The anisotropy at tree level is deferred to higher orders [at least $ \mathcal{O}(a^2) $] if $ c_t $, $ c_B $ and the coefficient of $ \mathcal{O}_{25} $ are chosen to be equal to $ \zeta $. %\newline
 Lastly,though $ \mathcal{O}(am_0) $~chiral corrections to bare parameters are prohibited by chiral symmetry, $ \mathcal{O}(a^2m_0^2) $~chiral corrections are required for the coefficient of the relevant counterterm.

\section{Symmetry of the fermion determinant}
\label{sec: Symmetry of the fermion determinant}

Invariance of the fermion determinant is inferred from invariance of the partition function, %\small
\begin{equation}
  \mathcal{Z}_{+\zeta} = N \int \mathcal{D}\bar\psi \mathcal{D}\psi\mathcal{D}U e^{-S^{f}_{+\zeta}[\psi,\bar\psi,U]-S^{g}[U]},
\end{equation}\normalsize
where $ S^{f}_{+\zeta}[\psi,\bar\psi,U] $ and $ S^{g}[U] $ are the fermion action of \mbox{eq.}~(\ref{eq: KW tl f action}) and Wilson's plaquette action including the three anisotropic counterterms and $ N $ is a normalisation constant. The fermions are formally integrated out and yield the fermion determinant and change the constant to $ N^\prime $, %\small
\begin{equation}
  \mathcal{Z}_{+\zeta} = N^\prime \int \mathcal{D}U \det(D^f_{+\zeta}[U])\ e^{-S^{g}[U]}.
  \label{eq: full qcd pf}
\end{equation}\normalsize
Next, the fermion determinant satisfies ($ C=i\gamma^0\gamma^2 $ is the charge conjugation matrix) %\small
\begin{equation}
  \det (D^{f}_{+\zeta}[U]) = \det (D^{f}_{+\zeta}[U] C^{-1} C) = \det (C D^{f}_{+\zeta}[U] C^{-1}) = \det (D^{f}_{-\zeta}[U^c])^T,
  \label{eq: cc of determinant}
\end{equation}
where gauge fields $ U $ are replaced by charge conjugated fields $ U^c $, the Wilczek parameter $ \zeta $ changes its sign and the Dirac operator is transposed. Since the gauge action and the measure are invariant under charge conjugation and the determinant is invariant under transposition, $ \mathcal{Z} $ satisfies
\begin{equation}
  \mathcal{Z}_{+\zeta} 
  = N^\prime \int \mathcal{D}U \det(D^f_{-\zeta}[U^c])^T\ e^{-S^{g}[U]} 
  = N^\prime \int \mathcal{D}U^c \det(D^f_{-\zeta}[U^c])\ e^{-S^{g}[U^c]}
  = \mathcal{Z}_{-\zeta}.
  \label{eq: full qcd pf w U^c}
\end{equation}\normalsize
After relabeling $ U^c $ to $ U $, it follows that the partition function is an even function of $ \zeta $. Since each of the transforms $ \widehat{C} $, $ \widehat{\Theta} $ or $ \widehat{T} $ flips the sign of $ \zeta $, the partition function is invariant under either. Due to the invariance of the gauge action and the measure under each, the fermion determinant is invariant as well. An even simpler proof uses the $ T_n $ of the unitary transform $ \widehat{T} $ [\mbox{cf. eq.}~(\ref{eq: unitary transform T})] instead of $ C $. Thus, the full QCD vacuum is invariant under $ \widehat{C} $, $ \widehat{\Theta} $ and $ \widehat{T} $ and has more symmetry than the action.

\section{ Time reflection of correlation functions}
\label{sec: Time reflection of correlation functions}

 Invariance of the Karsten-Wilczek action under either $ \widehat{C\Theta} $- or $ \widehat{T\,\Theta} $- transformations is reflected in invariance of the Hamiltonian $ \widehat{H} $. A generic correlation function $ \mathcal{C}(t_f-t_i) $ between initial and final states $ \widehat{O}_{i}^\dagger | \Omega \rangle $ and $ \widehat{O}_{f}^\dagger |\Omega \rangle $  using the transfer matrix $ \widehat{U}(t)=e^{-\widehat{H}t} $ reads \small
 \begin{align}
   \mathcal{C}(t_f-t_i) =&\ \langle \Omega | \widehat{O}_f e^{-\widehat{H}(t_f-t_i)} \widehat{O}_i^\dagger | \Omega \rangle,
 \end{align}\normalsize
 where $ |\Omega \rangle $ is the invariant vacuum. Invariance of the vacuum for full QCD is proved in section~\ref{sec: Symmetry of the fermion determinant}. In particular, $ \widehat{C} |\Omega \rangle = | \Omega \rangle = \widehat{\Theta} | \Omega \rangle $. Due to invariance of $ \widehat{H} $ under $ \widehat{C\Theta} $, one has $ \widehat{U}(t)  = \widehat{C}^\dagger \widehat{\Theta}^\dagger \widehat{U}(-t) \widehat{C} \widehat{\Theta} $:
 \small
  \begin{align}
   \mathcal{C}(t_f-t_i) =&\ \langle \Omega | \widehat{O}_f \widehat{C}^\dagger \widehat{\Theta}^\dagger e^{-\widehat{H}(t_i-t_f)} \widehat{\Theta} \widehat{C} \widehat{O}_i^\dagger | \Omega \rangle.
 \end{align}\normalsize
 If $ \widehat{O}_f $ and $ \widehat{O}_i $ both either commute or anticommute with $ \widehat{C} $, the operator $ \widehat{C} $ is moved past $ \widehat{O}_f $ and $ \widehat{O}_i $, \small
  \begin{align}
   \mathcal{C}(t_f-t_i) =&\ \langle \Omega | \widehat{C}^\dagger \widehat{O}_f \widehat{\Theta}^\dagger e^{-\widehat{H}(t_i-t_f)} \widehat{\Theta} \widehat{O}_i^\dagger \widehat{C} | \Omega \rangle,
 \end{align}\normalsize
 and leaves the vacuum invariant. Insertion of $ \mathbf{1} = \widehat{\Theta} \widehat{\Theta}^\dagger $ between vacuum states and operators yields\small
  \begin{align}
   \mathcal{C}(t_f-t_i) =&\ \langle \Omega | \widehat{\Theta}^\dagger \widehat{\Theta} \widehat{O}_f \widehat{\Theta}^\dagger e^{-\widehat{H}(t_i-t_f)} \widehat{\Theta} \widehat{O}_i^\dagger \widehat{\Theta}^\dagger \widehat{\Theta} | \Omega \rangle \\
    =&\ \langle \Omega | (\widehat{\Theta}\widehat{O}_f \widehat{\Theta}^\dagger) e^{-\widehat{H}(t_i-t_f)} (\widehat{\Theta} \widehat{O}_i^\dagger \widehat{\Theta}^\dagger) | \Omega \rangle = \widehat{\Theta} \mathcal{C}(t_f-t_i).
 \end{align}\normalsize
 This is manifest $ \widehat{\Theta} $~invariance of $ \widehat{C} $~invariant correlation functions. Since all dimension five operators explicitly break each of $ \widehat{C} $~, $ \widehat{\Theta} $~and $ \widehat{T} $~symmetries, odd powers of these operators must cancel if the correlation function has such symmetry. Therefore, leading discretisation effects are suppressed to $ \mathcal{O}(a^2) $ in $\widehat{C} $~invariant correlation functions. 
 Analogous results rely on $ \widehat{T} $ instead of $ \widehat{C} $ and require $ \widehat{O}_{i,f} $ that both either commute or anticommute with $ \widehat{T} $. As purely gluonic operators obviously commute with $ \widehat{T} $ (which affects only fermions), $ \widehat{\Theta} $~invariance and suppression of $ \mathcal{O}(a) $~corrections are apparent for gluonic observables. 
%  Even if operators $ \widehat{O}_{i,f} $ do not generate $ \widehat{C} $ or $ \widehat{T} $ eigenstates from the vacuum, these eigenstates are obtained from averaging with the correlation function using either $ \widehat{C}\widehat{O}_{i,f}\widehat{C} $ using charge conjguated gauge fields or $ \widehat{T}\widehat{O}_{i,f}\widehat{T} $. 
 \textbf{Caveat:} \textbf{Generic composite operators of fermion fields that lack additional symmetry under $ \widehat{C} $ or $ \widehat{T} $ may have $ \mathcal{O}(a) $~corrections.}

\section{  Decomposition of interacting fields} 
\label{sec: Decomposition of interacting fields}

 The free field has two components related by the unitary transform $ \widehat{T} $ of \mbox{eq.}~(\ref{eq: unitary transform T}), %\small
 \begin{align}
   \psi_n = \sum\limits_m g^\phi_{n,m}\phi_m + T_n g^\chi_{n,m} \chi_m, &\quad
   \bar{\psi}_n = \sum\limits_m \bar{\phi}_m g^{\phi\dagger}_{m,n}+ \bar{\chi}_m g^{\chi\dagger}_{m,n} T_n,
   \label{eq: free decomposition}
 \end{align}\normalsize
 where restriction of the components' four-momenta is realised through support on multiple sites. Two kernels $ g^\phi_{m,n} $, $ g^\chi_{m,n} $ (an example is found in~\cite{tib}), which satisfy $ g^\chi_{m,n} =T_m g^\phi_{m,n} T_n = \delta_{m,n}+\mathcal{O}(a) $ implement extended support. Contraction of unlike components generates terms with alternating sign in correlation functions. The same mechanism~\cite{alt} generates parity partner contributions for na\"{i}ve or staggered fermions. With interactions, the relevant operator of \mbox{eq.}~(\ref{eq: relevant term}) cancels divergent interaction effects that would spoil the continuum limit. In the following, approximate tuning is assumed -- a mismatch $ \delta c = c-c(g_0^2) $ of the parameter is supposed to be small. This mismatched relevant term would spoil the continuum limit of kernels like $ g^\phi_{m,n} $, $ g^\chi_{m,n} $.
 A local field transform %\small
 \begin{align}
   \psi_n \to \psi^c_n = \xi_n \psi_n, &\quad 
   \bar{\psi}_n \to \bar{\psi}^c_n= \bar{\psi}_n \xi_n^*, & \quad \xi_n \equiv e^{i\varphi n_0},\ \varphi \equiv \frac{\delta c}{1+d} 
 \end{align}\normalsize
 that modifies the temporal boundary condition (\mbox{cf.}~\cite{bbtw}) shifts the mismatch $ \delta c $ up to order $ \mathcal{O}(a,\varphi^2) $ into the definition of the fields $ \psi^c $, $ \bar{\psi}^c $. The lattice product rule that is applied requires linearisation in $ \varphi $ and $ a $. The field transform's parameter $ \varphi $ is constrained by the given parameters $ c $ (or rather $ \delta c $) and $ d $. As the relevant operator of \mbox{eq.}~(\ref{eq: relevant term}) anticommutes with $ \widehat{T} $, absorption of $ \delta c $ into two interacting components requires opposite phases [\mbox{cf.~eq.}~(\ref{eq: free decomposition})], % \small
 \begin{align}
   \psi_n =&\ \sum\limits_m \xi_n g^\phi_{n,m}[U] \phi_m^c + \xi_n^* T_n g^\chi_{n,m}[U] \chi_m^c, & \quad
   \bar{\psi}_n =&\ \sum\limits_m \bar{\phi}_m^c g^{\phi\dagger}_{m,n}[U] \xi_n^* + \bar{\chi}_m^c g^{\chi\dagger}_{m,n}[U] T_n \xi_n
,
 \end{align}\normalsize
 where kernels $ g^\phi_{m,n}[U] $, $ g^\chi_{m,n}[U] $ are outfitted with Wilson lines. This definition decomposes local fermionic bilinears at leading order into contributions with different space-time quantum numbers,%\small
 \begin{align}
   \bar{\psi}_n \mathcal{M} \psi_n =&\
   \xi_n^2 \bar{\chi}_k g^{\chi\dagger}_{k,n}[U] T_n\mathcal{M} g^{\phi}_{n,l}[U] \phi_l
   +\xi_n^{*2} \bar{\phi}_k g^{\phi\dagger}_{k,n}[U] \mathcal{M}T_n g^{\chi}_{n,l}[U] \chi_l 
   \nonumber \\
   +&\ \bar{\phi}_k g^{\phi\dagger}_{k,n}[U] \mathcal{M} g^{\phi}_{n,l}[U] \phi_l 
   +\bar{\chi}_k g^{\chi\dagger}_{k,n}[U] T_0 \mathcal{M}T_0 g^{\chi}_{n,l}[U] \chi_l.
   \label{eq: decomposition of bilinears}
 \end{align}\normalsize
  Dirac structures of bilinears with unlike (\mbox{e.g.} $ \bar{\phi} \mathcal{M}T_0 \chi $) or like (\mbox{e.g.} $ \bar{\phi} \mathcal{M} \phi $)  components differ by $ T_0 $. Phase factors $ (-e^{\pm 2i\varphi})^{n_0}=(-1)^{n_0}\xi_n^{(*)\,2} $ for unlike components in \mbox{eq.}~(\ref{eq: decomposition of bilinears}) generate oscillations in correlation functions with a frequency that depends on the mismatch $ \delta c $. Except for this $ \delta c $-dependence, this is well-known for na\"{i}ve and staggered fermions~\cite{alt}. If linearisation in $ \varphi $ and $ a $ and decomposition with kernels $ g^{\phi,\chi}_{n,m}[U] $ are applicable in the non-perturbative regime, the observed frequency is shifted by $ \omega_c = 2|\varphi| $ unless $ c $ is tuned properly. This prediction yields a new tuning condition that is put to the test in numerical simulations.

\section{ Pseudoscalar correlation functions}
\label{sec: Pseudoscalar correlation functions}

\begin{figure}[hbt]
 \begin{picture}(360,130)
   \put(010 , 0.0){\includegraphics[scale=0.80]{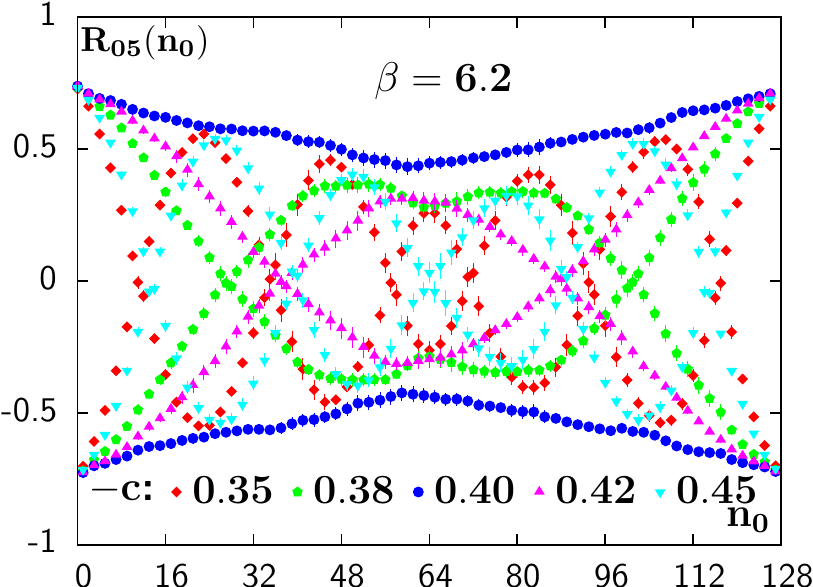}}
   \put(230.0, 0.0){\includegraphics[scale=0.80]{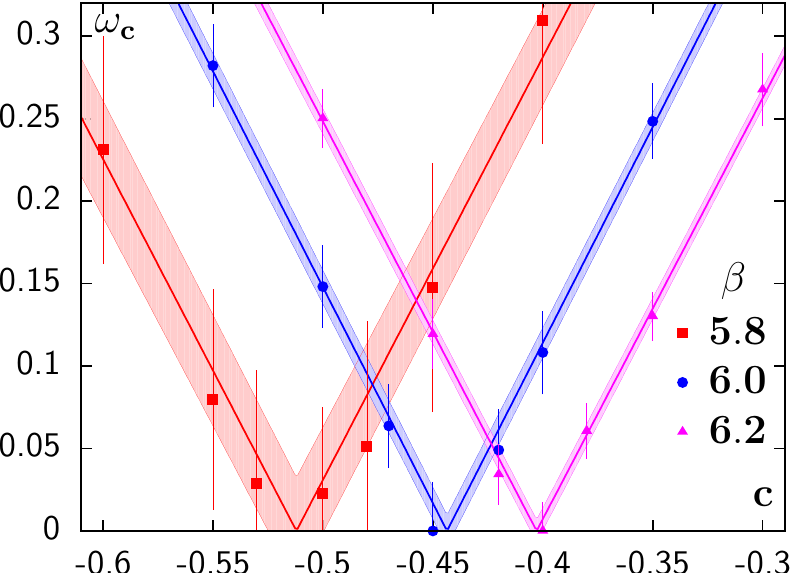}}
 \end{picture}
 \caption{Left: The frequency of oscillations in the ratio $ R_{05}(n_0) =\mathcal{C}_{00}(n_0)/\mathcal{C}_{55}(n_0) $ differs from $ \pi $ by a shift $ \omega_c $ that depends on $ c $. Right: The frequency shift $ \omega_c $ is linearly interpolated as $ \omega_c \propto | c-c_0 | $.}
 \label{fig: oscillations}
\end{figure}
\begin{figure}[hbt]
 \begin{picture}(360,130)
   \put(010.0 , 0.0){\includegraphics[scale=0.80]{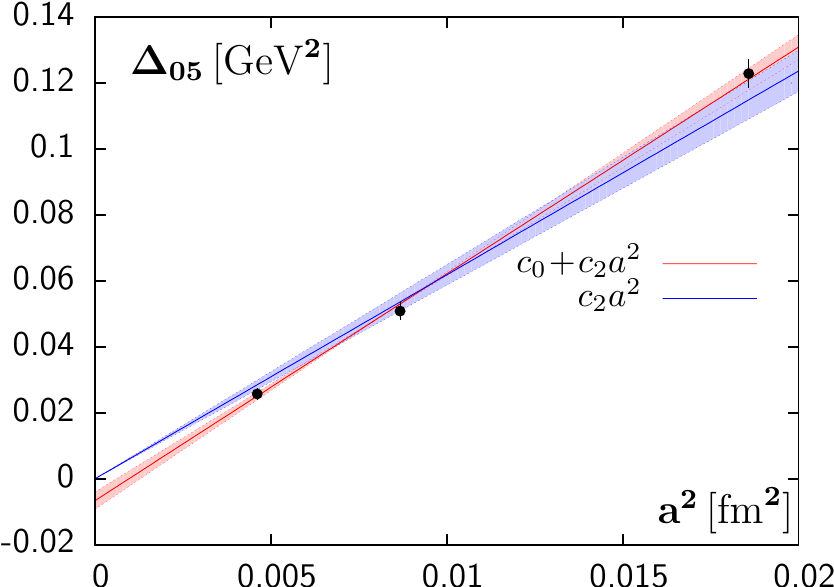}}
   \put(230.0 , 0.0){\includegraphics[scale=0.80]{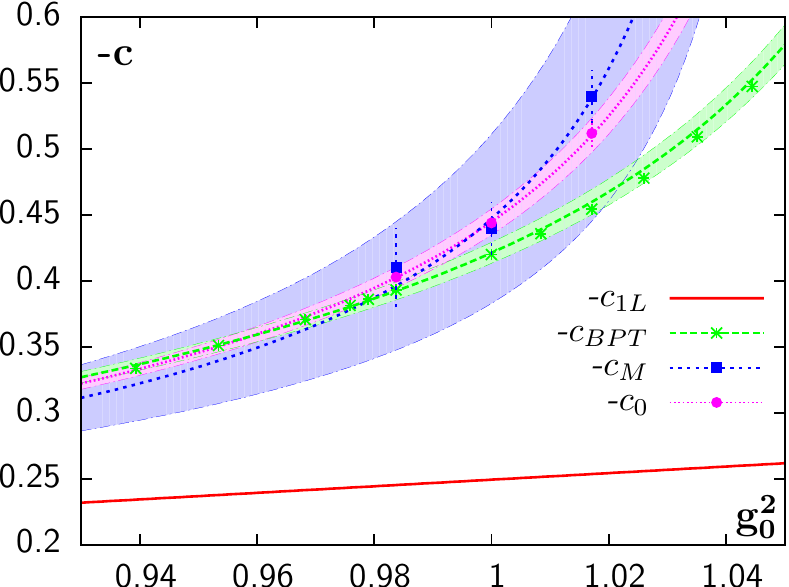}}
 \end{picture}
  \caption{Left: The splitting between pseudoscalar masses in $ \mathcal{C}_{00}(n_0) $ and $\mathcal{C}_{55}(n_0) $ vanishes as $ \mathcal{O}(a^2) $. Right: $ c_{1L} $ at one loop falls short non-perturbative results ($ c_M $, $ c_0 $). The figure is explained in the text.}
 \label{fig: continuum}
 \end{figure}

Pseudoscalars are the ground states of $ \gamma^0 $~and $ \gamma^5 $~channels due to \mbox{eq.}~(\ref{eq: decomposition of bilinears}). Oscillations are apparent in the ratio $ R_{05}(n_0) =\mathcal{C}_{00}(n_0)/\mathcal{C}_{55}(n_0) $ [\mbox{cf.} left plot in figure~\ref{fig: oscillations}].
 Good frequency resolution requires a long time direction. The frequency shift is studied on ten quenched configurations of $ 128\!\times\!24^3 $~lattices with pseudoscalar masses $ r_0 M_{55}\simeq 450\,\mathrm{MeV} $ and perturbatively tuned $ d $. Splitting between pseudoscalar masses in both channels, %\small
 \begin{align}
   \Delta_{05} = r_0^2(M_{00}^2-M_{55}^2),
 \end{align}\normalsize
 causes a small exponential decay that broadens the peaks in frequency spectra and causes a systematical error that dominates over the statistical error. The frequency is shifted as $ \omega_c \propto |c-c_0| = | \delta c | $ and agrees with the prediction of linearity [\mbox{cf.} right plot in figure~\ref{fig: oscillations}]. The error of $ c_0 $ is seen to diminish for finer lattices. %\newline
 The mass splitting $ \Delta_{05} $ is nearly constant over the range $ 250\,\mathrm{MeV} \lesssim r_0 M_{55} \lesssim 650\,\mathrm{MeV} $ and approaches the continuum limit as $ \mathcal{O}(a^2)$ [\mbox{cf.} left plot in figure~\ref{fig: continuum}]. $ 200(100) $ quenched configurations on $ 48\!\times\!24^3(32^3) $-lattices with $ \beta=5.8,6.0(6.2) $ are used. The oscillating ground state of $ \mathcal{C}_{00}(n_0) $ is degenerate with a Goldstone boson in the continuum limit. The ground state of $ \mathcal{C}_{55}(n_0) $ behaves as a Goldstone boson with quenched chiral logarithms at finite lattice spacing. These observations attest to the applicability of the decomposition of interacting fields. 

\section{ Non-perturbative tuning}

 Various methods for non-perturbative tuning of $ c $ are compared in the left plot in figure~\ref{fig: continuum}. Interpolations use a Pad\'{e} fit up to $ \mathcal{O}(g_0^4) $, where $ \mathcal{O}(g_0^2) $ is fixed as the one-loop result $ c_{1L} $ of \mbox{eq.}~(\ref{eq: 1L results}):
 \begin{align}
  c(g_0^2) = \tfrac{c_{1L}(g_0^2) + N\,g_0^4}{1+D\,g_0^2}.
 \end{align}
 The two fit parameters $ N $ and $ D $ for $ c_0 $ defined by $ \omega_c=0 $ read
 \begin{align}
   N_0 = -0.2163(8) &,\quad D_0 = -0.926(2),
 \end{align}
and agree within errors with $ c_M $ defined by minimal mass anisotropy~\cite{wcw}, though uncertainties of $ c_M $ are much larger. For fine lattices ($ a \leq 0.06\,\mathrm{fm} $), $ c_{BPT} $ from boosted perturbation theory~\cite{lm} is consistent with non-perturbative results. Thus, use of $ d_{BPT} $ appears justifiable. \newline

\textbf{Acknowledgements:} The speaker thanks \mbox{C. Seiwerth} for technical support and \mbox{P. Petreczky} for invaluable discussions.  
Simulations were performed on the cluster ``Lily'' at the Institute for Nuclear Physics, \mbox{Univ.} of Mainz. 
This work was supported by the ``Excellence Cluster Universe'' and by the Research center ``Elementary Forces \& Mathematical Foundations'' (EMG).

\enlargethispage{\baselineskip}

\end{document}